\documentclass{ar-1col-S2O}
\usepackage[numbers]{natbib}
\usepackage{url}
\usepackage{amsmath,amssymb}
\usepackage{braket}
\usepackage{multirow}

\usepackage[linktocpage]{hyperref}
\hypersetup{colorlinks=true,citecolor=blue,linkcolor=blue, urlcolor=blue, breaklinks=true}

\usepackage{xcolor}
\usepackage[mathscr]{euscript}
\usepackage[scr=boondox]{mathalpha}
\newcommand{\SO}{{\mathscr{S}_{\text{o}}}}

\newcommand{\Cmo}{\hat{C}_{M_{\text{o}}}}
\newcommand{\MO}{M_{\text{o}}}

\newcommand{\Geff}{G^{\text{eff}}}

\newcommand{\PO}{\vec{{\mathscr{P}}_{\text{o}}}}
\newcommand{\OO}{\text{o}}

\newcommand{\U}{\text{U}}

\newcommand{\Z}{{\mathbb Z}}

\jname{Xxxx. Xxx. Xxx. Xxx.}
\jvol{AA}
\jyear{YYYY}
\doi{10.1146/((please add article doi))}

\begin{document}

\markboth{Manjunath et al.}{Crystalline topology}

\title{Crystalline topological invariants in quantum many-body systems}

\author{Naren Manjunath$^1$ and Maissam Barkeshli$^2$ 
\affil{$^1$Perimeter Institute for Theoretical Physics, 31 Caroline St N, Waterloo, ON N2L 2Y5, Canada; email: nmanjunath@perimeterinstitute.ca}
\affil{$^2$Department of Physics and Joint Quantum Institute, University of Maryland, College Park, Maryland 20742, USA; email:maissam@umd.edu}}

\begin{abstract}
Crystalline symmetries give rise to topological invariants that can distinguish quantum phases of matter. Understanding these in strongly interacting systems is an ongoing research direction requiring non-perturbative methods. Recent developments have demonstrated that even classic models, like the Harper-Hofstadter model of free fermions on a lattice in a magnetic field, yield a host of crystalline symmetry protected topological invariants. Here we review some of these developments, focusing mainly on how to characterize, classify, and detect invariants arising from lattice translation and rotation symmetries along with charge conservation in two-dimensional systems, including integer and fractional Chern insulators. 
\end{abstract}

\maketitle

\tableofcontents

\section{INTRODUCTION}
Crystalline symmetry -- symmetry arising from lattice translations, rotations, and reflections -- underlies a remarkable number of phenomena in physics. From crystal structure and band theory to spontaneous symmetry breaking in charge density waves, magnetically ordered systems, and superconductivity, the crucial role of crystalline symmetry is widely appreciated. A more subtle effect of crystalline symmetry arises in topologically ordered states, like fractional Chern insulators and quantum spin liquids, where their very existence is forced by considerations of crystalline symmetry, namely mixed anomalies arising from filling constraints \cite{Savary2017, cheng2016lsm, else2020topological}. The latest chapter in this century-long saga is in understanding more subtle quantum effects centered on how crystalline symmetries give rise to topological invariants in symmetry-preserving quantum phases of matter, and how these are manifested in observable properties.

Historically, the issue of invariants of quantum phases of matter arising from crystalline symmetry came from two main lines of work. The first stretches back to the theory of quantum spin liquids, where symmetry-enriched topological orders were studied and began to be classified using gauge mean-field theory constructions \cite{wen04,wen2002psg}. The second started with the discovery of time-reversal invariant topological insulators \cite{kane2005a,kane2005b}, which gave way to the systematic study of topological band theory and where crystalline topological insulators soon became a central topic \cite{hasan2010,qi2010RMP}. However these approaches were missing fundamental ingredients. For instance, topological band theory is of limited use in understanding quantum phases of matter where interactions can play a pronounced role \cite{fidkowski2010effects,wang2014,Senthil2015SPT}. 

The last fifteen years has seen major advances in a general theory of topological invariants with symmetry in interacting quantum many-body systems. These include the idea of symmetry-protected topological phases and symmetry-enriched topological phases, along with the theoretical frameworks to classify and characterize them, culminating in approaches based on group cohomology \cite{dijkgraaf1990,Chen2013SPT}, cobordism theory \cite{kapustin2014SPTbeyond,kapustin2014b} and invertible topological quantum field theories (TQFT) \cite{freed2016}, G-crossed braided tensor categories (BTCs) \cite{Barkeshli2019, Barkeshli2022SPT, bulmashSymmFrac,bulmash2022anomaly,aasen2021characterization}, and rigorous operator algebraic methods \cite{ogata2021,sopenko2021}. The case of crystalline symmetry is of particular interest given its relevance to  realistic systems, and brings with it additional complications relative to the case of internal symmetries. 

\begin{figure}[t]
\includegraphics[width=5in]{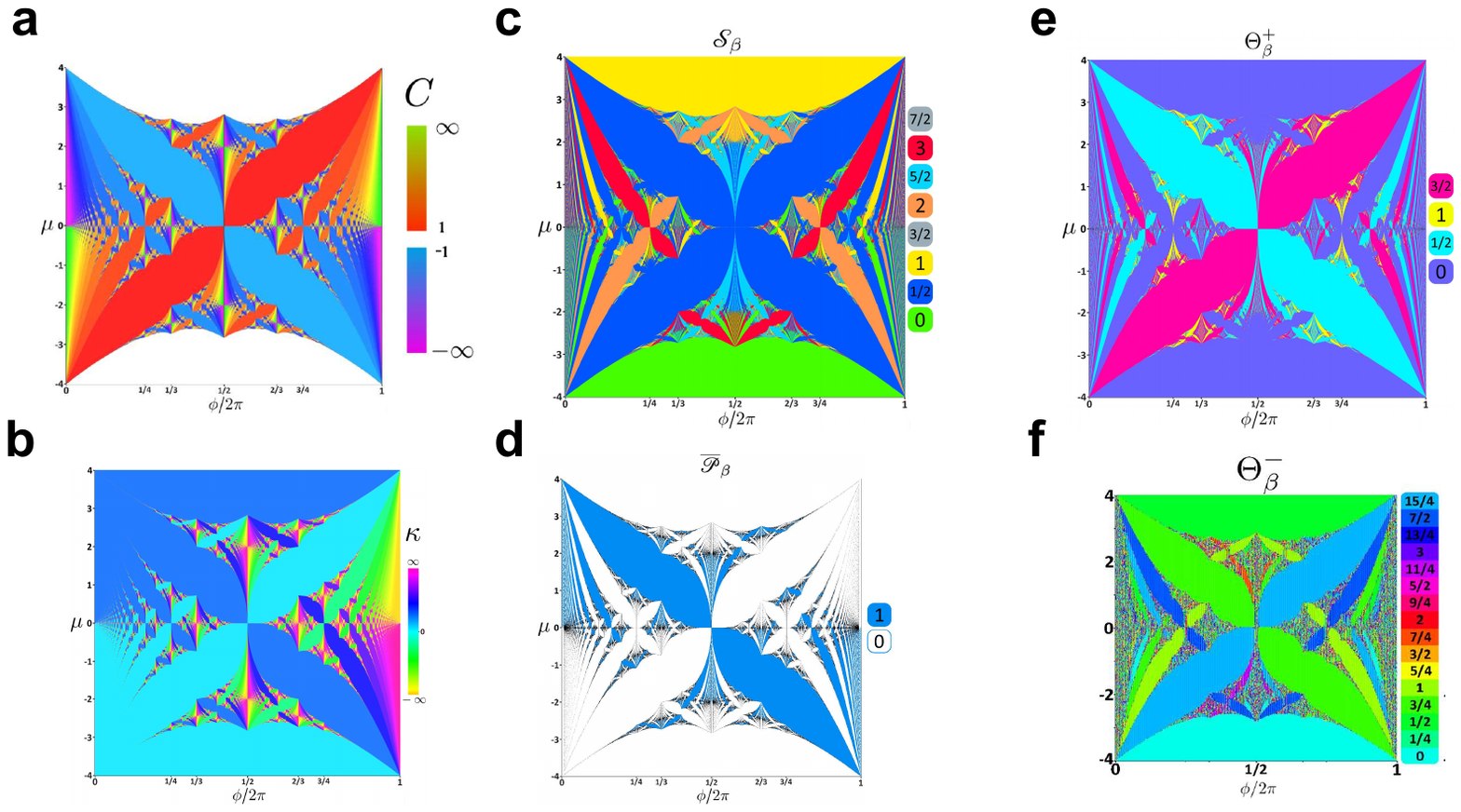}
\caption{Hofstadter butterflies for the invariants (a) $C$, Chern number; (b) $\kappa$, related to the filling; (c) $\mathscr{S}_{\beta}$, discrete shift; (d) $\vec{\mathscr{P}}_{\beta}$, electric polarization; (e) $\Theta^+_{\beta}$ and (f) $\Theta^-_{\beta}$, both partial rotation invariants, where $\beta$ is a vertex.}
\label{Figure_1}
\end{figure}

In this article, we review some developments involving crystalline symmetry protected invariants that arise in well-known quantum systems. After surveying some general methods used to attack these problems, we mainly focus on a physically relevant case of two-dimensional systems of fermions, with $U(1)$ charge conservation symmetry along with lattice translation and rotational symmetry. Recently a complete classification and characterization was provided in this case, uncovering new invariants and several different ways of measuring them. These include new crystalline invariants in integer and fractional Chern insulators. A highlight of these results was the first discovery in over forty years -- since the quantized Hall conductivity derived by TKNN \cite{thouless1982} -- of new topological invariants arising in the Harper-Hofstadter model of fermions hopping on a lattice in a magnetic field. This has led to a variety of ways to color Hofstadter's famous butterfly with new topological invariants arising from crystalline symmetry \cite{zhang2022fractional,zhang2022pol,zhang2023complete}, as displayed in Fig. \ref{Figure_1}.

\subsection{Problem statement} The theoretical study of topological phases of matter with symmetry is often focused on two related but separate questions, that of \textit{classification} and \textit{characterization}. The classification problem involves identifying the number of distinct topological phases in a given symmetry class and spatial dimension, together with any group structure involved in the problem, such as by stacking different phases together. The characterization problem is far richer: what are the appropriate theoretical frameworks to capture topological properties of strongly interacting quantum many-body systems? Given a microscopic Hamiltonian or ground state wave function, how do we extract the topological invariants? What manifestation do the invariants have in physical observables?

\subsection{Assumptions}

\subsubsection{Degrees of freedom} The most fundamental assumption is about whether the local degrees of freedom are bosons or fermions; the theory of fermionic systems has several additional complications relative to the fermionic case. Since free fermions can still be non-trivial, it is important to consider both free and interacting fermionic systems separately. From a field-theoretic perspective, gapped phases of fermions have an emergent anomalous 1-form symmetry, corresponding to the fermion worldline, which significantly enriches the problem \cite{gaiotto2014,Gaiotto:2015zta,tata2023anomalies,Barkeshli2022SPT,bulmashSymmFrac,bulmash2022anomaly,aasen2021characterization}.

\subsubsection{Symmetries} We assume the ground state of the system preserves a symmetry group $G$. In the fermionic case, there is a symmetry group $G_f$, referred to as the fermionic symmetry group. $G_f$ includes an order-2 subgroup, denoted $\mathbb{Z}_2^f$, generated by fermion parity $(-1)^F$. The bosonic symmetry group, which is the group that acts non-trivially on bosonic operators, is a quotient, $G_b = G_f/\mathbb{Z}_2^f$. $G_f$ is a central extension of $G_b$ by $\mathbb{Z}_2^f$. To fully define the symmetry, we need to specify which group elements are anti-unitary and how they act on space, which can be done by defining group homomorphisms from the symmetry group into $\mathbb{Z}_2$ and the isometries of space, respectively. In this review, symmetries always act on the full system (0-form symmetries).

\subsubsection{Topological order} In this review we focus on ground states of gapped Hamiltonians, which have exponentially decaying correlations. It is useful in setting up the problem to distinguish between two broad classes, {\it invertible} and {\it non-invertible} phases of matter. A gapped ground state $\ket{\Psi}$ is said to be invertible (sometimes referred to as short-range entangled) if there exists an `inverse' state $\ket{\Psi^{-1}}$, such that $\ket{\Psi} \otimes \ket{\Psi^{-1}}$ can be adiabatically connected to a trivial product state. An invertible state is degenerate on any closed manifold, and does not have topological excitations with fractional braiding statistics (anyons). Furthermore, if $\ket{\Psi}$ and $\ket{\Psi'}$ are two invertible states, then so is $\ket{\Psi} \otimes \ket{\Psi'}$; this implies that the classification forms a group under the tensor product or `stacking' operation. Any state that can arise in free fermion systems is invertible, such as integer quantum Hall states, Chern insulators, Majorana nanowires, topological insulators~\cite{hasan2010} and superconductors~\cite{bernevig2013topological} (as well as their `higher-order' or `crystalline' versions \cite{ando2015topological}). 

\begin{marginnote}
\entry{SPT}{Symmetry-protected topological state: a special case of an invertible state which can be connected to a trivial product state upon breaking the protecting symmetry.}
\entry{HO}{Higher order: a prefix applied to topological states protected by crystalline symmetries.}
\entry{CI}{Chern insulator: An invertible state with $\U(1)$ charge conservation and crystalline symmetries, and a nonzero quantized Hall conductance.}
\end{marginnote}

A special class of invertible phases are symmetry-protected topological (SPT) states~\cite{Senthil2015SPT}. These are states that can be adiabatically connected to the identity when the symmetry is broken, without even requiring an inverse state. For example, time-reversal invariant topological insulators are SPTs, while the IQH and Majorana nanowire are invertible states but not SPTs. 

In contrast, a non-invertible state, sometimes said to have \it intrinsic topological order \rm and \it long-range entanglement\rm, has a ground state degeneracy on a torus and topological excitations with fractional braiding statistics (anyons). Notable examples include the fractional quantum Hall states, fractional Chern insulators, fractional topological insulators, and (gapped) quantum spin liquids.  

\begin{marginnote}
\entry{FCI}{A topologically ordered state with $\U(1)$ charge conservation and crystalline symmetries, and a nonzero quantized Hall conductance.}
\entry{QSL}{A topologically ordered state with $\mathrm{SO}(3)$ spin rotation and crystalline symmetries}
\end{marginnote}

\subsection{Overview of approaches} 

\begin{figure}[t]
\includegraphics[width=5in]{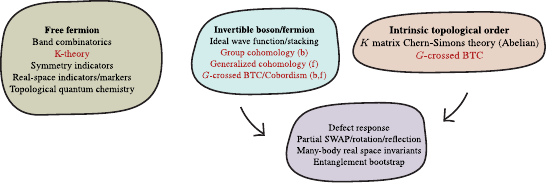}
\caption{Summary of approaches to classify and characterize crystalline topological states. Mathematical frameworks for classification are shown in red; (b), (f) denote bosonic and fermionic classifications.}
\label{Figure_2}
\end{figure}

A summary of the various approaches for classification and characterization in different contexts is given in Figure~\ref{Figure_2}.

\subsubsection{Free-fermion states}

Topological phases of translationally invariant free (non-interacting) fermions can be described by topological band theory, which falls naturally into the mathematical framework of $K$-theory and its generalizations. The seminal early results on classifying free fermion topological states with internal symmetries (the `periodic table' of topological insulators and superconductors) \cite{kitaev2009,ryu2010,Chiu2016review} were later generalized to crystalline symmetries in a series of works \cite{benalcazar2014classification,Shiozaki2014TCIS,Kruthoff2017bandCombinatorics,Po2017symmind,bradlyn2017topological,Shiozaki2018AH,shiozaki2022AHSS}. Each crystalline topological phase is specified by a set of integers $\{n^{\bf k}_R\}$, where $n^{\bf k}_R$ denotes the number of bands that transform under the irreducible representation (irrep) $R$ of the site-symmetry group fixing the momentum ${\bf k}$ in the Brillouin zone. The choices of $n_{R}^{\bf k}$ at different ${\bf k}$ satisfy various compatibility conditions. Note that certain classifications~\cite{Po2017symmind,bradlyn2017topological} place all atomic insulators- states representable in terms of exponentially localized Wannier orbitals- in the trivial phase; however, atomic insulators with non-zero filling are non-trivial in the many-body classifications discussed below, as they cannot necessarily be adiabatically connected to each other while preserving the symmetries.

The linear combinations of $n_R^{\bf k}$ that detect free fermion states beyond atomic insulators can be identified using approaches such as `symmetry indicators' \cite{Po2017symmind,watanabe2018structure,tang2019comprehensive} `topological quantum chemistry' (TQC) \cite{bradlyn2017topological,Cano_2021,Elcoro2021tqc} and $K$-theory~\cite{shiozaki2022AHSS,ono2024AHSS}. Other approaches extract these invariants from dislocation, disclination and corner charges \cite{Miert2018dislocationCharge,Benalcazar2019HOTI,Li2020disc} or other real-space indices~\cite{velury2025marker,hwang2025RSI}, and are closer in spirit to those we will discuss below for interacting systems. Finally, free-fermion systems in the presence of magnetic flux $\phi$ per unit cell have magnetic translation symmetry and the added complexity of \textit{projective} representations at site-symmetry groups; recent works have explored band representations~\cite{herzog2020Hofstadter}, real-space indices~\cite{herzog2022Hofstadter} and a generalization of TQC~\cite{Fang2023SI} to this setting, but a comprehensive treatment for $\phi \neq 0$ is still lacking. 

\subsubsection{Invertible states with interactions} 
\noindent \textbf{Ideal wave functions}. One set of classification strategies relies on constructing a simple representative ground state wave function for each crystalline topological phase, assuming a suitable ideal limit \cite{song2017,Huang2017,Song2018TopoCrystals,Cheng2018fSPT,Song2020,zhang2022realspace}. In these constructions, a crystalline SPT wave function is constructed as a tensor product of lower-dimensional states defined at the high symmetry points and lines of the real-space unit cell. These constructions generally assume a `Wannier limit' in which the degrees of freedom are fully supported at such high symmetry regions, or argue why any representative state in this phase can be deformed systematically into such a limit without undergoing a phase transition \cite{Huang2017}. The various SPT invariants can then be understood in terms of the symmetry eigenvalues or quantum numbers of the localized degrees of freedom. The classification can be obtained by counting the number of distinct values these quantum numbers can take, subject to certain equivalences. Although this approach gives a simple and intuitive picture of the ground state, it does not explain how these topological invariants should be detected far away from the idealized limit, when there is no neat interpretation in terms of localized symmetry eigenvalues. 

\noindent\textbf{Topological quantum field theory (TQFT)}. Quantized responses can be described in terms of topological field theory. For instance, the quantized Hall conductivity can be understood in terms of a Chern-Simons Lagrangian with a $U(1)$ gauge field:
\begin{align}
    \mathcal{L} = \frac{C}{4\pi} A \wedge d A ,
\end{align}
where $A$ is the $U(1)$ gauge field and $C$ is the Chern number. From a more general perspective, this topological field theory can be understood through the framework of group cohomology. Given a symmetry group $G$, the cohomology group $\mathcal{H}^{d+1}(G, U(1))$ classifies topological terms that can appear in a Lagrangian in $d+1$ space-time dimensions for a $G$ background gauge field \cite{dijkgraaf1990,Chen2013SPT}. The quantized Hall conductivity can be understood through this lens of group cohomology by setting $G = U(1)$. This more general perspective allows us to understand the case with discrete symmetries as well. The topologically invariant Chern-Simons Lagrangian for a background $G$ gauge field $A$ allows us to construct an \it invertible \rm TQFT: the path integral $\mathcal{Z}[M^{d+1}, A] = e^{i \int_{M^{d+1}} \mathcal{L}[A]}$ is a $U(1)$ phase on every $d+1$-dimensional space-time manifold. There are invertible TQFTs that cannot be described by group cohomology and require more powerful frameworks, such as cobordism theory \cite{kapustin2014SPTbeyond,kapustin2014b, freed2016, kapustin2015fSPT}.

The case of crystalline symmetries raises additional complications. We can make progress by observing that in many-body systems, crystalline symmetries act in the infrared (IR) low energy continuum quantum field theory via a combination of internal symmetries and spatial symmetries of the continuum space-time. Therefore, a path to understanding topological phases with a crystalline symmetry group $G$ is to understand all possible ways of enriching the low-energy quantum field theory with an internal symmetry $G$. This suggests the following general conjecture: \textit{The classification of topological states for a spatial symmetry $G$ (or $G_f$) and given additional symmetry data is in one-to-one correspondence with the classification of topological states for an effective internal symmetry $\Geff$ (or $G_f^{\text{eff}}$) and additional effective symmetry data.} This was referred to as the ``crystalline equivalence principle" (CEP) in ~\cite{Thorngren2018,Else2019}. The relationship between $G^\text{eff}$ and $G$ is particularly complicated in the fermionic case; concrete formulas to express the effective internal symmetry data in terms of the spatial symmetry data were given in Reference~\citenum{manjunath2022mzm}.

Although the CEP does not have a completely general mathematical justification, it has been verified in many explicit examples \cite{Song2020,zhang2022realspace,manjunath2022mzm} and there are additional formal justifications covering various special cases~\cite{debray2021invertible}. The peculiarities in the fermionic case can be understood as a simple consequence  \cite{barkeshli2026disclinations} of the fact that fermions in TQFT necessarily have spin-1/2 under space-time rotation symmetries due to the spin-statistics theorem.

\noindent \textbf{Symmetry defect response}. The TQFT approach assumes that gapped phases of matter can be fully characterized by TQFTs, which assign a topologically invariant path integral to every space-time manifold. But what is the justification for this assumption? A more physical approach to characterizing symmety-protected topological invariants of gapped systems is to analyze the properties of symmetry defects, which can always be done on some simply connected patch of space. The electric charge bound to magnetic flux is one example of this, were the magnetic flux can be thought of as a symmetry defect for $U(1)$ symmetry. A comprehensive theory of symmetry defects in two spatial dimensions requires the mathematical framework of $G$-crossed braided tensor categories \cite{Barkeshli2019}, which captures their universal braiding and fusion propoerties. The TQFT can then be directly defined mathematically from the categorical description of the symmetry defects \cite{walker2006,barkeshli2019tr,Bulmash2020absolute,tata2023anomalies,walker2021universal}. 

In the case of crystalline symmetries, we need to understand the response of the system to crystalline symmetry defects. For translational and rotational symmetry, these are lattice dislocations and disclinations. It is fruitful to model these lattice defects in a TQFT description using ``crystalline gauge fields," which are background gauge fields whose fluxes keep track of lattice defects. In the case of CIs and FCIs, one can construct a topological response theory involving crystalline gauge fields, generalizing the one for the quantized Hall conductivity~\cite{manjunath2021cgt,Manjunath2020fqh}. As we review below, this theory predicts additional topological invariants, such as a `discrete shift' and a quantized electric polarization, and predicts that they can be characterized by the fractional charge bound to lattice disclinations and dislocations respectively~\cite{Liu2019ShiftIns,zhang2022fractional,zhang2022pol}. Such defect-based approaches give an important partial characterization of many-body topological invariants.

\noindent \textbf{Partial symmetries:} 
Ground state expectation values of partial symmetry operations -- symmetries applied to a subregion of the system -- can be used to extract a variety of invariants. For example, ground state expectation values of suitable SWAPs or spatial rotations can detect central charges~\cite{Qi2012momentumpolarization, FQHEDMRG, kobayashi2023extracting} and internal symmetry SPT invariants~\cite{Zaletel2014bosonicSPT,shiozaki2017matrix,turzillo2025}. In these cases, TQFT can provide a systematic way to identify ground state expectation values that extract non-trivial topological invariants, by relating them to partition functions on appropriate manifolds. In the crystalline case, the TQFT identifies the right ground state expectation values to be partial rotations and reflections \cite{shiozaki2017invt,zhang2023complete,herzogarbeitman2022interacting,Shiozaki2018antiunitary,kobayashi2024FCI}, which we will discuss further in Sec.~\ref{sec:inv-prot}. Partial symmetries can in many cases give a \textit{complete} characterization of invertible as well as topologically ordered states with crystalline symmetry~\cite{zhang2023complete,kobayashi2024FCI,calvera2025wallpaper}. Furthermore, partial symmetries extract topological invariants from a single ground state without any defect insertions, adding to a line of work on extracting topological invariants from single bulk ground states~\cite{levin2006,kitaev2006topological,shiozaki2017invt,dehghani2021,cian2021,cian2022extracting,Kim2022ccc,fan2022}.

The classification of invertible states with free and interacting fermions can be very different~\cite{fidkowski2010effects,gu2014,Morimoto2015}, as illustrated in Table~\ref{tab:classification}: certain free fermion states can be trivialized upon adding suitable interactions, while certain interacting states do not have any free-fermion analog. It is therefore useful to construct a `free-to-interacting map' relating the two classifications. Examples of this map have been constructed for internal~\cite{Chen2019freeinteracting} as well as crystalline symmetries~\cite{zhang2022realspace,manjunath2023characterization,lee2026freeinteracting}.

\begin{table}[t]
    \centering
    \caption{Classification and overcomplete invariants for free and interacting invertible fermionic states with representative symmetries. Parantheses $\{\}$ mean that we should consider all possible high-symmetry points ${\bf p}$ (momentum space) or $\OO$ (real space). 
    }
    \label{tab:classification}
    \begin{tabular}{l||l|l||l|p{0.3\textwidth}} \hline
         & \multicolumn{2}{c||}{\textbf{Free fermion}} &   \multicolumn{2}{c}{\textbf{Interacting fermion}} \\ \hline 
        Symmetry & Classification & Invariants & Classification & Invariants \\
        $\U(1)^f \times \Z_4$ & $\Z^5$ & $C,n^{{\bf \Gamma}}_q$ &$\Z^2 \times \Z_8 \times \Z_2$ & $c_-,C;\SO,\ell_{\OO}$ or $\Theta^{\pm}_{\OO}$ \\
        $\U(1)^f \times \text{p4}$ & $\Z^9$ & $C,\{n^{{\bf p}}_q\}$ &$\Z^3 \times \Z_8 \times \Z_4^2\times\Z_2$ & $c_-,C,\nu;\{\SO,\ell_{\OO}\}$ or $\{\Theta^{\pm}_{\OO}\}$  \\ \hline
    \end{tabular}
\end{table}

\subsubsection{Topologically ordered states} Non-invertible states, with intrinsic topological order, in the absence of symmetry are characterized by a unitary modular tensor category (UMTC) $\mathcal{C}$ \cite{moore1989,witten1989,moore1991,Kitaev2006,wang2008,nayak2008}. $\mathcal{C}$ captures the anyons, their fusion rules, and a set of $F$ and $R$ symbols specifying additional associativity and braiding properties. A topologically ordered state with additional global symmetry is referred to as a symmetry-enriched topological (SET) state. Bosonic SETs in (2+1) dimensions are characterized by $G$-crossed braided tensor categories \cite{Barkeshli2019,Barkeshli2020Anomaly}, and these have been generalized to fermionic SETs \cite{bulmashSymmFrac,bulmash2022anomaly,aasen2021characterization}. In the case where all anyons are Abelian (implying that fusing anyons has a unique outcome), these classifications reduce to the earlier framework of $K$ matrix Chern-Simons theory \cite{Wen1995,lu2012,Lu2016}. The classification in the bosonic case can roughly be summarized in terms of three levels of data: (i) anyon permutations under the global symmetry; (ii) fractional symmetry quantum numbers carried by the anyons (known as \textit{symmetry fractionalization}); and (iii) the additional freedom to stack an SPT state, which changes the properties of symmetry defects but not those of the anyons.

Prior classifications of crystalline SET phases have studied symmetry fractionalization in spin liquids~\cite{Essin2013SF,zaletel2017,ye2024QSL}, as well as complete classifications for FQH states and FCIs \cite{manjunath2021cgt,Manjunath2020fqh}. Many of the characterization methods applicable to invertible states naturally carries over to the case of SETs, and enables the detection of symmetry fractionalization. Ideal wave functions of FCIs obtained from the parton construction have recently been characterized using both partial symmetries \cite{kobayashi2024FCI} and defect response~\cite{zhang2025FCI}.

\section{Crystalline invariants in invertible topological states}\label{sec:invertible}

\subsection{Preliminaries}\label{sec:prelim}
 
In this review, we will for illustration restrict to specific unitary, orientation-preserving symmetries of the square lattice, which we define as $G_f = \U(1)^f \times_{\phi} \mathrm{p4}$ when considering fermionic states, and $G = \U(1) \times_{\phi} \mathrm{p4}$ when considering bosonic states. Here p4 is the orientation-preserving square lattice wallpaper group; its real-space unit cell is shown in \textbf{Figure 1}. The superscript on $\U(1)^f$ means that the fermion parity appears as a subgroup. The subscript $\times_\phi$ means that translations generated by $\tilde{T}_{\bf x},\tilde{T}_{\bf y}$ do not commute; instead, we have
\begin{equation}
    \tilde{T}^{-1}_{\bf y} \tilde{T}^{-1}_{\bf x} \tilde{T}_{\bf y} \tilde{T}_{\bf x} = e^{i \phi \hat{N}},
\end{equation}
where $\hat{N}$ is the fermion number operator. The high-symmetry points in the p4 unit cell are $\alpha,\beta$ (order 4) and $\gamma$ (order 2). We can canonically choose a rotation operator of order $M_{\OO}$ at $\OO$ denoted $\Cmo^+$, satisfying $(\Cmo^+)^{\MO} = +1$. It will be convenient to define a second rotation operator $\Cmo^- := e^{\frac{i \pi}{\MO} \hat{N}} \Cmo^+$, which satisfies $(\Cmo^-)^{\MO} = (-1)^F$.

Our main numerical example when discussing invertible states is the classic non-interacting Hofstadter model \cite{hofstadter1976} on the square lattice. It has the Hamiltonian
\begin{equation}
    H = -\mu \sum_i c^{\dagger}_i c_i-t \sum_{<ij>} c^{\dagger}_i e^{-i A_{ij}}c_j  + \operatorname{ h.c.},
\end{equation}
where the vector potential $A_{ij}$ assigns flux $\phi$ per unit cell. This Hamiltonian has the above symmetry $G_f$, and was the first model in which the quantized Hall conductance was computed, by TKNN in 1982 \cite{thouless1982} (see Figure~\ref{Figure_1} (a)). The filling $\nu$ satisfies 
\begin{equation}\label{eq:nu}
    \nu = C\frac{\phi}{2\pi} + \kappa~,
\end{equation}
where $\kappa$ is a second integer invariant (see Figure~\ref{Figure_1}(b). The Hofstadter model has been experimentally realized in moir{\'e} superlattice systems \cite{Dean2013,Hunt2013hb,Saito2021,Eric2018moire}, ultracold atoms \cite{aidelsburger2013,miyake2013,kennedy2015}, and photonics \cite{hafezi2013,ozawa2019}. Although the model is non-interacting, we will use it to illustrate several characterization schemes that also generalize to many-body invertible states and topologically ordered states. Generalizing to other orientation-preserving wallpaper groups, namely p1, p2, p3 and p6, is relatively straightforward~\cite{zhang2022pol,zhang2023complete,manjunath2023characterization}. 

In the rest of this section, we will discuss the classification from both ideal wave function and TQFT perspectives. We will then use the response theory from TQFT to identify charge and angular momentum responses, which extract new quantized invariants we denote as $\SO$ (the discrete shift), $\PO$ (a quantized electric polarization) and $\ell_{\OO}$, as depicted in Figure~\ref{Figure_1}.

\subsection{Classification for a single rotation center}\label{sec:inv-single}
We first consider the far simpler case of invertible states with symmetry $\U(1)^f \times \Z_4$, where $\Z_4$ denotes a fourfold rotation symmetry about a rotation center $\OO$ (which we will also fix as the origin of space) with rotation operator $\hat{C}_{4,\OO}$. In Sec.~\ref{sec:p4-classif} and subsequently, we generalize to the case $\U(1)^f \times_{\phi} \text{p4}$, with multiple rotation centers.

\subsubsection{Invariants without crystalline symmetry} Invertible states without any symmetry are characterized by the chiral central charge $c_-$, which $c_-$ sets the thermal Hall conductance, $\kappa_H = \frac{\pi c_-T}{6}$. $c_-$ can take integer or half-integer values, and in free fermion states, $2c_-$ corresponds to the Chern number of their associated Bogoliubov- de Gennes (BdG) Hamiltonian. The $\U(1)^f$ symmetry endows invertible states with a second integer invariant, the Chern number $C$, which sets the (electrical) Hall conductance, $\sigma_H = C/2\pi$ in natural units. While $C = c_-$ for free fermions, the TQFT approach (see below) shows that in general
\begin{equation}\label{eq:C_vs_c}
    C = c_- \mod 8~.
\end{equation}
A representative state with $C = 1, c_-=1$ is the usual IQH state with its lowest Landau level filled. A representative state with $C=8, c_-=0$ is the bosonic IQH state~\cite{senthil2013,he2015biqh}. A representative of each topological phase can be obtained by stacking copies of the above two states with one that satisfies $C = c_-=0$, which we now discuss.

\subsubsection{Ideal wave function constructions}

The `Wannier limit' is particularly simple in the case with a single rotation center: the only degrees of freedom are localized at $\OO$, while all others are pushed away to infinity. Each Wannier limit state is labelled by a set of quantum numbers associated with the degrees of freedom at $\OO$, and different labellings can be equivalent under symmetric moves. For example, in the case of bosonic SPT states, the wave function is labelled by a pair $(N_{\OO},m_{\OO}) \in \Z \times \Z_4$ corresponding to the charge and angular momentum of bosons at $\OO$. Now a rotationally symmetric orbit of 4 bosons carries zero angular momentum, and can be symmetrically moved to infinity, giving an equivalence $N_{\OO} \simeq N_{\OO} + 4$. The overall classification of bosonic SPT states in the Wannier limit with $C=0$ is therefore given by pairs $(N_{\OO} \mod 4, m_{\OO} \mod 4) \in \Z_4 \times \Z_4$. Similar arguments in the case of invertible fermionic states with $G_f = \U(1)^f \times \Z_4$ lead to a labelling of Wannier limit states (with $C = c_- = 0$) as $(n_{\OO},m_{\OO}) \in \Z \times \Z_4$. However, a rotationally symmetric orbit of four fermions carries charge 4 and \textit{non-trivial} angular momentum 2 mod 4, which is a consequence of the fermionic anticommutation relations; this implies $
    (n_{\OO},m_{\OO}) \simeq (n_{\OO}-4,m_{\OO}+2)~$~\cite{cheng2018rotation,manjunath2023characterization}.
After incorporating this equivalence, the only invariant quantities are $n_{\OO} + 2 m_{\OO} \mod 8$ and $m_{\OO} \mod 2$. Therefore, ideal wave functions with symmetry $G_f$ and $c_- = C=0$ are classified by the group $\Z_8 \times \Z_2$ \cite{Cheng2018fSPT}. Note that the charge and angular momentum numbers are not independent in the fermionic case.

\subsection{TQFT approach}\label{sec:inv-tqft}

Deriving the TQFT action involves a number of technical subtleties, especially in the fermionic case. Although a full discussion of these issues is outside the scope of this review, we will summarize the main steps involved before stating and justifying the result.
 
\subsubsection{Crystalline equivalence principle} In the present case, the CEP states that if the physical rotation operator satisfies $(\Cmo^+)^{\MO} = +1$, then the effective internal $\Z_{\MO}$ generator ${\bf h}$ must satisfy ${\bf h}^{\MO} = (-1)^F$, and vice versa. This has been shown to give the correct classification in many examples by matching to independent classifications based on ideal wave functions \cite{Song2020,zhang2022realspace,debray2021invertible,manjunath2022mzm}. Note that the microscopic symmetry $G_{\text{UV}}$ embeds into the symmetry $G_{\text{IR}}$ of the low-energy theory via a group homomorphism $\rho: G_{\text{UV}} \rightarrow G_{\text{IR}}$. The CEP is really a statement about how spatial operators in $G_{\text{UV}}$ act in the IR theory, where they decompose into a space-time part and an internal part. A simple justification of the CEP can be made by considering a system whose low-energy theory is a 2-component Dirac fermion ${\bf \Psi}(t,\vec{x})$~\cite{barkeshli2026disclinations}. In this case the operators $\Cmo^{\pm}$ take
\begin{equation}
    {\bf \Psi}(t,\vec{x}) \rightarrow  \U_{\MO}^{\pm}\Lambda(2\pi/\MO){\bf \Psi}(t,\vec{x}')
\end{equation}
where $\Lambda$ is a space-time rotation and $\U_{\MO}^{\pm}$ are internal $\U(2)$ rotations. Crucially, the Dirac fermion transforms as a spinor under space-time rotations, therefore $(\Lambda(2\pi/\MO))^{\MO} = (-1)^F$; and to ensure that the overall space-time operation has the correct group structure, the internal rotations must satisfy $(\U_{\MO}^{\pm})^{\MO} = (\Cmo^{\pm})^{\MO} \times (-1)^F$. In particular, if rotations in the UV define a group $\Z_{\MO}$ where $\MO$ is even, then the effective internal rotation defines a group $\Z_{2\MO}^f$, and vice versa. The group structure does not change when $\MO$ is odd.

\subsubsection{Gauge fields}Next, we define gauge fields for the effective internal symmetry. A $G_f$ gauge field on a 3-dimensional space-time manifold $\mathcal{M}^3$ is a pair $(A,\omega)$ where $A,\omega$ are real-valued differential 1-forms on $\mathcal{M}^3$. $A$ is the usual vector potential. The flux $dA$ (where $d$ denotes the exterior derivative) corresponds to magnetic flux. $\omega$ is a gauge field for an effective $\Z_8^f$ internal symmetry\footnote{In the literature, $\omega$ is sometimes defined as a $\Z_4$ gauge field~\cite{zhang2022fractional,manjunath2023characterization}, but those works derived the action after defining $\omega$ as a gauge field for $G_b$. Here we have defined $\omega$ as a $G_f$ gauge field from the start, choosing our conventions to give the same quantization of invariants as the above works.} and charge under $\omega$ is physically interpreted as angular momentum under $\Cmo^+$. We choose conventions so that $d\omega$ has the same quantization as the allowed fluxes of a discrete rotational symmetry, physically corresponding to the disclination angle $\Omega$ of lattice disclinations. To do so, we constrain  its holonomy over any closed curve $C$: $\oint_{C} \omega \in \frac{2\pi}{4} \Z$. We further choose the following flux quantization convention on a closed 2-manifold $\mathcal{M}^2$:
\begin{equation}
    \int_{\mathcal{M}^2} dA \in 2\pi \Z; \quad \int_{\mathcal{M}^2} d\omega \in 4\pi \Z~. 
\end{equation}
The condition on $\omega$ reflects the fact that the curvature (or total disclination angle) of any closed orientable manifold $\mathcal{M}^2$ is $4\pi (1-g)$ where $g$ is the genus; it is a consequence of the $\Z_8^f$ group structure of the effective internal symmetry coming from the CEP.

Note that we formally derive the TQFT by defining the gauge fields as simplicial cochains on a triangulated manifold, but we state the result here in terms of real-valued differential forms because they are more familiar in the condensed matter literature. A dictionary relating the simplicial and continuum formulations is given in References~\cite{manjunath2021cgt,zhang2022fractional}.

\subsubsection{Construction of effective action} Finally, we construct the general TQFT action in terms of $(A,\omega)$. The general construction for an invertible state with symmetry $G_f$ and $c_- \neq 0$ was discussed in Reference~\citenum{zhang2022fractional} using technical results from References~\citenum{barkeshli2021invertible,aasen2021characterization}. Here we will briefly survey the technical arguments.

In the case of bosonic SPT states, the effective action $\mathcal{L}$ is a Dijkgraaf-Witten term~\cite{dijkgraaf1990} taking the form $\mathcal{L} = B^* \nu_3$ where $B$ denotes the $G_b$ gauge field\footnote{Formally, $B$ is a map from $\mathcal{M}^3 \rightarrow BG_b$ where $BG_b$ is the classifying space of $G_b$. Here $*$ denotes the pullback from the classifying space.}. $\nu_3$ is a representative cocycle in the cohomology group $\mathcal{H}^3(G_b,\U(1))$. For $G_b = \U(1) \times \Z_4$, this gives the clasification $\Z \times \Z_4^2$, which agrees with the ideal construction assuming $C=0$.

Deriving the analogous action for invertible \textit{fermionic} states requires the machinery of higher category theory and $G$-crossed BTCs~\cite{barkeshli2021invertible,aasen2021characterization,ning2021enforced}. The strategy in Reference~\citenum{barkeshli2021invertible} is to gauge the fermion parity symmetry, resulting in a topological order classified by Kitaev's `16-fold way' (and specified by $2c_-$ mod 16)~\cite{Kitaev2006} enriched by $G_b$ symmetry. This is described mathematically by a $G_b$-crossed BTC, which encodes the fusion and braiding properties of anyons and symmetry defects through a number of consistency conditions and equivalences. One then classifies the different solutions of the $G_b$-crossed consistency equations modulo certain redundancies which must also be meaningful in the original invertible state. These results can also be stated in terms of $\text{Spin}(2c_-)$ Chern-Simons theory~\cite{barkeshli2021invertible}.

The resulting classification can be compactly expressed as follows. Each invertible fermionic state with symmetry $G_f$ is specified by the \textit{data} $\{c_-,n_1,n_2,\nu_3\}$, where 
\begin{equation}
 n_1 \in \mathcal{H}^1(G_b,\Z_2)~; \quad n_2 \in C^2(G_b,\Z_2)~; \quad \nu_3 \in C^3(G_b,\U(1))~.
\end{equation}
(Here $C^n(G_b,M)$ is the group of $n$-cochains, or maps from $G_b^n  \rightarrow M$ where $G_b^n$ is the direct product of $n$ copies of $G_b$.) The homomorphism $n_1$ specifies the presence of unpaired Majorana zero modes at symmetry defects; the 2-cochain $n_2$ specifies the quantum numbers of fermion parity fluxes; and the 3-cochain $\nu_3$ specifies the quantum numbers of $G_b$ symmetry defects. The data must satisfy the \textit{consistency conditions}
\begin{equation}
    dn_1 = 0~; \quad dn_2 = \mathcal{O}_3[\omega_2,c_-,n_1] \mod 2~; \quad d\nu_3 = \mathcal{O}_4[\omega_2,c_-,n_1,n_2] \mod 1~,
\end{equation}
where $[\omega_2] \in \mathcal{H}^2(G_b,\Z_2)$ specifies the group extension of $G_b$ by $\Z_2^f$ that defines $G_f$. $\mathcal{O}_3, \mathcal{O}_4$ are \textit{obstructions} that must be cohomologically trivial to admit any solutions. Finally, there are \textit{equivalences}
\begin{equation}
    n_2 \simeq n_2 + \omega_2~; \quad \nu_3 \simeq \nu_3 + \mathcal{B}_3[\omega_2]~,
\end{equation}
coming from relabelling fermion parity fluxes and $G_b$ symmetry defects, respectively, by fermions. In the case of spatial symmetries, one must replace $\omega_2$ by the data $\omega_2^{\text{eff}}$ associated to $G_f^{\text{eff}}$. Expressions for $\mathcal{O}_3, \mathcal{O}_4, \mathcal{B}_3$ are derived in Reference~\cite{barkeshli2021invertible}.

An effective action can be obtained from these data using the relation $L = B^*\nu_3$, where $B$ is a $G_b$ gauge field, only if $c_-$ is an integer and there are no unpaired Majorana zero modes at $G_b$ symmetry defects, i.e. $n_1 = 0$. Both conditions hold when the system has $\U(1)^f$ symmetry~\cite{manjunath2022mzm}. Below, we will state the final result in terms of the $G_f$ gauge field $(A,\omega)$, which is more natural and was defined above.

\subsubsection{Result and interpretation} The above calculation gives the following result:

\begin{align}\label{eq:mainresponse-rot}
    \mathcal{L} &= \frac{C}{4\pi} A \wedge dA + \frac{\SO}{2\pi} A \wedge d\omega +\frac{\ell_{\OO}}{4\pi} \omega \wedge d\omega  ~,
\end{align}
where $\wedge$ denotes the wedge product of differential forms, and

\begin{align}\label{eq:quantization_S_l}
    C &= c_- + 8 k_1~;\quad
    \SO = \frac{c_-}{2} + [s_{\OO}]_2 + 2 k_{2,\OO} \mod 8~; \quad
    \ell_{\OO} = \frac{c_-}{4} + 2k_{3,\OO} \mod 8~.
\end{align}
$k_{1},k_{2,\OO},k_{3,\OO}$ are `integration constants' and physically correspond to bosonic SPT states that can be stacked without altering the quantum numbers of fermions. The quantity $s_{\OO}$ is an `intrinsically fermionic' contribution which must be zero in any bosonic SPT. There is an additional subtle contribution coming from the framing anomaly which is written in terms of the space-time spin connection~\cite{manjunath2021cgt}. The equivalence on $\nu_3$ implies that $(k_{2,\OO},k_{3,\OO}) \simeq (k_{2,\OO} + 2, k_{3,\OO} + 2)$, where all other coefficients remain unchanged. Finally, the independent integer-quantized invariants are
\begin{align}\label{eq:FTresult_C4}
    c_-,k_1 \in \Z~;\quad 
    I_1 := \SO- \ell_{\OO}-\frac{c_-}{4} \in \Z_8 ~;\quad 
    I_2 := \frac{1}{2}\left(\ell_{\OO} - \frac{c_-}{4}\right) \in \Z_2~.
\end{align}
Thus the full classification is $\Z^2 \times (\Z_8 \times \Z_2)$, and the last two factors agree with the ideal wave function approach, which assumed $C = c_-=0$.

The first term of Eq.~\ref{eq:mainresponse-rot} captures the Chern number $C$; note that Eq.~\ref{eq:C_vs_c} comes out as a derivation. The second term with coefficient $\mathscr{S}_{\OO}$ has a clear interpretation: in a region with flux $d\omega$ corresponding to a disclination of angle $\Omega$, the `excess charge' is
$\int \frac{\delta \mathcal{L}}{\delta A} = \SO\frac{\Omega}{2\pi} \mod 1$. Below, we explain how to precisely define the `excess charge' in a microscopic model. Although the invariant $\ell_{\OO}$ apparently does not have a similar interpretation in terms of disclination defects, we will discuss a different way to characterize it in Sec.~\ref{sec:inv-prot}.

When $c_-=0$, $I_1$ and $I_2$ have the same quantization as $n_{\OO} + 2 m_{\OO} \mod 8$ and $m_{\OO} \mod 2$ from the ideal construction. This suggests the correspondence $\ell_{\OO} \simeq 2 m_{\OO} \mod 4$ and $\SO \simeq n_{\OO} \mod 4$.  

\subsection{Charge response from discrete shift}\label{sec:inv-shift}

\subsubsection{Quantization and main properties}Let us now study the second term with $\SO$ in detail and explain its quantization in Eq.~\ref{eq:quantization_S_l}.
\begin{enumerate}
    \item $2\SO \in \Z$. The condition $\int_{S^2} d\omega = 4\pi$ implies that the term with $\SO$ assigns a charge $2\SO$ on a sphere (or the surface of a cube), which must be an integer.
    \item $\SO \simeq \SO + 4$. Note that $\int_W d\omega = \oint_{\partial W} \omega$ is quantized in multiples of $\pi/2$ for any region $W$ with boundary. The charge $Q_W$ in a region containing an elementary disclination therefore receives a contribution $\SO/4$. However, only the fractional part of $Q_W$ is topologically invariant, since one can always attach a fermion to the disclination core and change $Q_W$ by an integer. Therefore, $\SO/4 \simeq \SO/4 + 1$.
    
    Eq.~\ref{eq:mainresponse-rot} predicts that the total charge in a region $W$ surrounding a disclination with disclination angle $\Omega$ is 

\begin{align}\label{eq:QW}
    Q_W &= C  \int_W \frac{dA}{2\pi} + \SO \int_W \frac{d\omega}{2\pi} \mod 1 ~.
\end{align}
Therefore, the theory splits the charge inside $W$ into two contributions, coming from the total flux in $W$ and from the discrete shift. In practice, however, it is highly non-trivial to correctly separate these contributions. The issue is that $\int_W dA$ measures the \textit{total} flux in $W$, including a background flux and an excess flux; and the background flux depends on subtle details of how the disclination is constructed, also on the choice of $\OO$. These subtleties have been carefully addressed in prior work~\cite{zhang2022fractional,zhang2022pol}; below we will illustrate the main steps for a special case that does not suffer from these subtleties.

\item \textit{In the Wannier limit, we have $\SO = n_{\OO}$. More generally, when $C = c_- = 0$, $\SO$ is an integer.} The claim $\SO = n_{\OO}$ can be tested by considering a cube with six p4 unit cells whose vertices are at the desired orign $\OO$; say $\OO = \alpha$ in this case. Then, if we set $W$ to be the entire cube in Eq.~\ref{eq:QW-final}, the total charge is $
        Q_W = 8 n_{\alpha} + 6 n_{\beta} + 12 n_{\gamma} = 6 \nu + 2\mathscr{S}_{\alpha}.$ Using $\nu = n_{\alpha} + n_{\beta} + 2 n_{\gamma}$, we immediately find that $\mathscr{S}_{\alpha} = n_{\alpha}$, as claimed. Finally, the quantization of $\SO$ should be robust throughout the gapped phase of matter, since the topological action is not affected upon perturbing away from the Wannier limit.
\item \textit{We have the important constraint} $   \SO = C/2 \mod 1$. The discrete shift is a crystalline analog of the Wen-Zee shift~\cite{Wen1992shift} in continuum quantum Hall states, which is detected by measuring the Hall viscosity \cite{Avron1995hvisc,Read2009hvisc,Read2011hvisc,haldane2009hall,haldane2011fqh,Gromov2014,Bradlyn2015,Gromov2015,schine2016,wu2017fqh}. There are two major differences between them: (i) the Wen-Zee shift is protected by continuous $\mathrm{SO}(2)$ rotational symmetry, and has a $\Z$ classification, in contrast to what we obtained above; (ii) the continuum shift is origin-independent, while the discrete shift $\SO$ fundamentally depends on a choice of rotation center ($\SO, \mathscr{S}_{\OO'}$ are generally distinct invariants). One can realize IQH states on the lattice (for example, by going to the $\phi \rightarrow 0$ limit of the Hofstadter model), which have the effective action
\begin{equation}
    \mathcal{L}_{\text{IQH}} = \sum_{n=0}^{C-1} \frac{1}{4\pi}(A + s_n \omega) d(A+ s_n \omega) + \mathcal{L}_{\text{anom}}~,
\end{equation}
where $s_n = n+ 1/2$ is the orbital angular momentum of the fermion in the $n$th Landau level (LL); this can be obtained, for example, by solving the Landau level problem on a sphere~\cite{Wen1992shift}. In other words, the $n$th Landau level has $\SO = n+1/2 = 1/2 \mod 1$. Stacking any $C$ of them gives $\SO = C/2 \mod 1$ in the Landau level limit. But a general state with symmetry $G_f$ is a stack of such Landau level states, together with some $C=0$ state, in which we showed that $\SO$ is an integer.  
\end{enumerate}

\begin{figure}[t]
\includegraphics[width=5in]{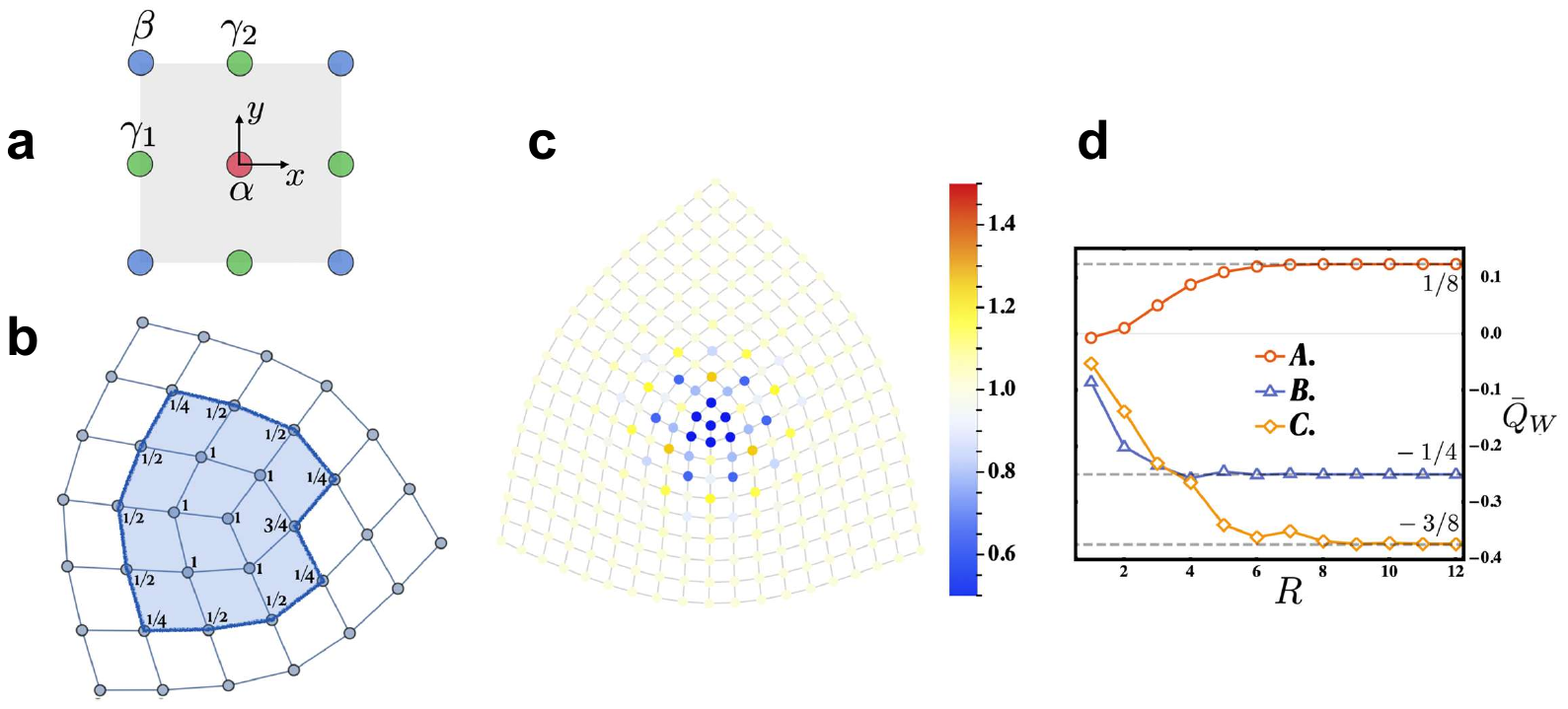}
\caption{Measurement of $\SO$. (a) Real-space unit cell of p4; (b) Weights for charge measurement; (c) Sample numerical plot of $Q_i$; (d) quantization of $\SO/4$ for different Hofstadter ground states as the radius $R$ of $W$ is increased.}
\label{Figure_3}
\end{figure}

\subsubsection{Numerical detection} Below we will consider the simplest situation of a disclination at a vertex, assuming uniform flux in each plaquette, and that the boundary of $W$ coincides with plaquette boundaries (see for example Figure~\ref{Figure_3}(b)). These assumptions ensure that and the number of unit cells in $W$, denoted $N_W$, is an integer. Under these assumptions, $\int_W dA = \phi N_W$. We also have $\int_W d\omega = \Omega$, while $\nu = C\phi/2\pi$ mod 1 from Eq.~\ref{eq:nu}. Combining these relations, we can recast Eq.~\ref{eq:QW} as follows:
\begin{equation}\label{eq:QW-final}
    Q_W= \nu N_W + \SO \frac{\Omega}{2\pi} \mod 1~.
\end{equation}
(In the general case, $N_W$ also depends on $\OO$; however, the full right-hand side of Eq.~\ref{eq:QW-final} is always origin-independent.) In a lattice model, we can compute $Q_W$ by assigning weights $\operatorname{wt}(v) = 1$ to each vertex $v$ in the interior of $W$, and $\operatorname{wt}(v) = \theta/2\pi$ if the interior of $W$ subtends an angle $\theta$ at $v$. These weightings ensure that $Q_W$ is additive on neighboring regions that share boundaries, which is a requirement of the field theory. Then, we have $Q_W = \sum_{v} \operatorname{wt}(v) Q_v$, where $Q_v$ is the ground state charge at $v$. Knowing $\nu, N_W$ and $\Omega$, we can extract $\SO$ from Eq.~\ref{eq:QW-final}. Note that although $\SO$ does not depend on translation symmetry, this method implicitly assumes a lattice model with translation symmetry. Some numerical results for the Hofstadter model are shown in Figure~\ref{Figure_3}. In several sample ground states, we see that the fractional excess charge quickly approaches a fixed constant value, enabling an unmbiguous determination of $\SO$. We can repeat this for a number of different lobes of the Hofstadter model, and extrapolate the result throughout the Hofstadter butterfly by fitting to an analytical formula~\cite{zhang2022fractional}. This finally gives the coloring shown in Figure~\ref{Figure_1}(c). The quantization is precisely as predicted above. Furthermore, $\SO$ is independent of the integer invariants $c_-, C, \kappa$, as demonstrated by two lobes with $c_- = C = 4, \kappa = -1$ that have different values of $\SO$. This demonstrates that $\SO$ is indeed a new invariant of the Hofstadter model.

\subsubsection{Origin-dependence, dual interpretation, and disorder} The origin-dependence of $\SO$ implies that extracting it at different origins $\OO$ gives additional information. In particular, taking appropriate differences of $\SO$ for different $\OO$ extracts a quantized electric polarization, which we discuss in Sec.~\ref{sec:pol}. Secondly, the term with $\SO$ in Eq.~\ref{eq:mainresponse-rot} has a dual interpretation: a ground state with excess magnetic field $\delta \Phi$ has excess angular momentum $\SO \delta\Phi/2\pi$. This dual interpretation has also been verified~\cite{You2020hoe,zhang2022fractional}, and is a very non-trivial check on the validity of the TQFT approach. Finally, upon adding weak disorder which does not close the gap, the values of $\SO$ in individual disorder realizations are not constant; however, their mean agrees with the value at zero disorder strength, and the variance in $\SO$ increases with disorder strength~\cite{zhang2022fractional}.

\subsection{Angular momentum response from partial rotations}\label{sec:inv-prot}

The following quantization of $\ell_{\OO}$ can be deduced following similar arguments as for $\SO$: (i) $4\ell_{\OO} \in \Z$; (ii) $\ell_{\OO} \simeq \ell_{\OO} + 2$; (iii) $\ell_{\OO} = C/4 \mod 1$.

Now although disclination charge gives a direct physical probe of $\SO$, the angular momentum response $\ell_{\OO}$ does not depend on $\U(1)^f$ symmetry, and cannot be detected through charge probes. Furthermore, it does not appear that $\ell_{\OO}$ can be similarly characterized using disclination defects. Since Eq.~\ref{eq:mainresponse-rot} does not directly motivate an observable, we will first need to make a claim and then justify it from different angles. 

\subsubsection{Main result} Recall the two rotation operators $\Cmo^{\pm}$, satisfying $(\Cmo^+)^{\MO} = +1$ and $(\Cmo^-)^{\MO} = (-1)^F$. Now, Eq.~\ref{eq:mainresponse-rot} is really a response theory for  rotations associated to $\Cmo^+$; therefore we will denote the coefficients of Eq.~\ref{eq:mainresponse-rot} as $\SO^+, \ell_{\OO}^+$. It will be useful to distinguish these from $\SO^-, \ell_{\OO}^-$, which are coefficients in a different response theory for the same state obtained by replacing $A \rightarrow A + \omega/2$ in Eq.~\ref{eq:mainresponse-rot}. Now the claim is that $\ell_{\OO}^{\pm}$ can be obtained from a `partial rotation', defined as follows. Consider an $M_{\OO}$-fold rotation center $\OO$, and define a rotationally symmetric region $D$ centered at $\OO$. Let $\Cmo^{\pm}|_D$ be the restriction of $\Cmo^{\pm}$ to $D$. Then, we claim that to leading order
\begin{align}\label{eq:ThetaO}
    \bra{\Psi}{\Cmo^{\pm}|_D}\ket{\Psi} \simeq   \exp\left({-\gamma_{\OO}^{\pm }|\partial D|}\right) \exp\left({ \frac{2\pi i}{M_{\OO}} K_{\OO}^{\pm}}\right) (1 + O(e^{- \epsilon |\partial D|}))~,
\end{align}
where the amplitude decreases with the size of the boundary $\partial D$, and the topologically invariant quantities are given by 
\begin{align}\label{eq:Theta_ModRed}
    \Theta_{\OO}^+ :=
        K^+_{\OO} \mod \frac{\MO}{2}, \;\;\;\; \Theta_{\OO}^- := K^-_{\OO} \mod \MO.
\end{align}
Furthermore, on the square lattice $\Theta^{\pm}_{\OO}$ are related to $\ell^{\pm}_{\OO}$ as follows:
\begin{align}\label{eq:ells_vs_Theta}
    \ell^{\pm}_{\OO} &=\begin{cases}
        \frac{11 \mp 1}{8}C+2\Theta^{\pm}_{\OO} \mod 4 \quad \OO=\alpha,\beta\\
        \frac{1\mp1}{4}C+2\Theta^{\pm}_{\OO} \mod 2\quad \OO=\gamma .
    \end{cases}
\end{align}

\subsubsection{Justification} An initial sanity check is to evaluate this expectation value when $\ket{\Psi_{\text{WL}}}$ is a Wannier limit state with charge $n_{\OO}$ and angular momentum $m_{\OO}$ at the rotation center. In this case, by picking $D$ to enclose the single point $\OO$ we trivially find that 
\begin{equation}
    \bra{\Psi_{\text{WL}}}{\Cmo^+|_D}\ket{\Psi_{\text{WL}}} = e^{i \frac{2\pi}{M_{\OO}}m_{\OO}}~, \quad \bra{\Psi_{\text{WL}}}{\Cmo^-|_D}\ket{\Psi_{\text{WL}}} = e^{i \frac{\pi}{M_{\OO}}(n_{\OO} + 2m_{\OO})}~.
\end{equation}
When $\MO = 4$, the partial rotation extracts $m_{\OO}$ mod 4 and $n_{\OO} + 2 m_{\OO}$ mod 8. The equivalence operation of moving four fermions to infinity takes $m_{\OO} \rightarrow m_{\OO} + 2$ but leaves $n_{\OO} + 2 m_{\OO}$ mod 8 invariant, justifying the modular reduction in Eq.~\ref{eq:Theta_ModRed}.

A more careful justification of the robustness of this observable is based on relating it to a TQFT partition function evaluated on a lens space $L(M_{\OO},1)$~\cite{Tantiv2017,shiozaki2017invt,turzillo2025}. Viewed in spacetime, the ground states $\ket{\Psi}$ and $\bra{\Psi}$ on a 2-sphere $S^2$ correspond to the TQFT partition function on two 3-dimensional disks $D^3$. The above expectation value corresponds to gluing the two copies of $D^3$ with a $2\pi/M_{\OO}$ rotation (and, in the case of $\Cmo^-$, an additional $\U(1)^f$ rotation). The manifold realized by this surgery is the lens space $L(M_{\OO};1)$, and it has fundamental group $\Z_{M_{\OO}}$ (it is the classifying space of $\Z_{M_{\OO}}$). The TQFT partition function on this manifold (and therefore the partial rotation) can take $M_{\OO}$ different values for a fixed $c_-$, which implies that this operator can distinguish the correct number of invertible states.
   
A still more powerful justification uses the assumption that the low-lying entanglement spectrum of the density matrix $\rho_D$ matches that of the CFT living on the boundary of $D$:
\begin{equation}
    \rho_D \simeq \rho_{CFT}.
\end{equation}
This assumption was first stated by Li and Haldane~\cite{Haldane2008entanglement}, and has been verified in a number of integer and fractional quantum Hall states, although there are known counterexamples. In this case, $\Cmo^{\pm}|_{\partial D}$ acts a translation by $|\partial D|/M_{\OO}$ in the CFT. Therefore, we can express the order parameter as the expectation value of a translation in the CFT, combined with some $\U(1)^f$ rotation:
\begin{align}\label{eq:Prot-main}
\bra{\Psi}\Cmo^{\pm}|_D\ket{\Psi} = \frac{\mathrm{Tr}[e^{iQ_{\MO}\frac{\pi}{\MO}}e^{i\tilde{P}\frac{|\partial D|}{\MO}}e^{-\frac{\xi}{v} H}]}{\mathrm{Tr}[e^{-\frac{\xi}{v} H}]}
\end{align}
where $\xi$ is the bulk correlation length, and $v$ is the velocity of excitations in the CFT; these set an effective inverse temperature $\beta = \xi/v$. $Q_{\MO}$ generates a subgroup of $\U(1)^f$ which is determined using the CEP. Eq.~\ref{eq:Prot-main} gives a remarkably precise formula to relate the order parameter to $\ell_{\OO}^{\pm}$~\cite{zhang2023complete}. The traces in Eq.~\ref{eq:Prot-main} involve a weighted sum over all the symmetry sectors of the CFT (which are labelled by the fermion and the fermion parity fluxes), and the weights depend on their rotational quantum numbers, which are fixed by the underlying TQFT, Eq.~\ref{eq:mainresponse-rot}. A detailed calculation~\cite{zhang2023complete} approximates the order parameter to leading order in terms of the $G$-crossed modular $T$ matrix of the TQFT \cite{Barkeshli2019}, with the result 
\begin{align}
\bra{\Psi}\Cmo^{\pm}|_D\ket{\Psi}_{\text{CFT}} &\propto 
e^{-\frac{2\pi i}{24} v_{\pm}(\MO)c_- + \frac{2\pi i}{M_{\OO}} \frac{\ell_{\OO}^{\pm}}{2}}  = e^{\frac{2\pi i}{\MO}K_{\OO}^{\pm}}.
\label{eq:mainresultCFT}
\end{align}
Here $v_{\pm}(\MO)$ are known constants. Eq.~\ref{eq:mainresultCFT} is in exact agreement with Eq.~\ref{eq:ThetaO}. It also allows us to relate $\ell_{\OO}^{\pm}$ (which appears in the modular $T$ matrix) to $\Theta^{\pm}_{\OO}$, finally giving Eq.~\ref{eq:ells_vs_Theta}.

\subsubsection{Numerical verification}

We show the final result for the partial rotations in the Hofstadter model in Figure~\ref{Figure_1}(e,f). Note that we numerically find $\bra{\Psi} \Cmo^+|_D \ket{\Psi} \sim e^{i \frac{2\pi}{\MO}K_{\OO}^+}$, where $K_{\OO}^+$ actually jumps by $2$ within a single lobe. Therefore, the topologically invariant part of the partial rotation is indeed $\Theta^+_{\OO} = K_{\OO}^+ \mod 2$, precisely as predicted by theory. On the other hand, $K^-_{\OO} \equiv \Theta^-_{\OO}$ \textit{directly} gives a $\Z_8$ invariant (Figure~\ref{Figure_1}(f)), and there are no jumps within a single lobe. This is again explained by the equivalences in the ideal wave function limit: when we increase $D$ to include orbits of 4 fermions, the total eigenvalue under $\Cmo^-$ does not change. Remarkably, $\Theta_{\OO}^{+}$ and $\Theta_{\OO}^-$ therefore give a complete characterization of topological invariants with $G_f = \U(1)^f \times \Z_4$, assuming $C, c_-$ are known. This also implies that they give an independent way to measure the discrete shift $\SO$. It can be shown~\cite{zhang2023complete} that
\begin{equation}
    2(\Theta^-_{\OO}-\Theta^+_{\OO}) = \mathscr{S}^+_{\OO} \mod \MO.
\end{equation}

\subsection{Classification with p4 spatial symmetry}\label{sec:p4-classif}
If we now consider the complete space group p4, with multiple rotation centers, the Wannier limit corresponds to a state in which the degrees of freedom are localized at $\alpha,\beta,\gamma$ (see Figure~\ref{Figure_3}(a)), and the associated invariants are $(n_{\OO},m_{\OO}), \OO = \alpha,\beta,\gamma$. The equivalences now permit degrees of freedom to be symmetrically moved between $\alpha$ and $\beta$, or $\alpha$ and $\gamma$. 

The bosonic SPT classification obtained this way is $\Z_4^3 \times \Z_2^2$, which agrees with the cohomology group $\mathcal{H}^3(G_b,\U(1))$ up to a $\Z$ factor coming from the Chern number~\cite{Huang2017}. In the fermionic case, the classification forms the group $\Z \times \Z_8 \times \Z_2 \times\Z_4^2$ when $C = c_- = 0$~\cite{manjunath2023characterization}. $C$ and $c_-$ give two additional integer invariants. The resulting classification $\Z^3\times \Z_8 \times \Z_2 \times\Z_4^2$ can be reproduced by the $G$-crossed BTC approach~\cite{manjunath2023characterization}.

The TQFT is now written in terms of a gauge field for the group $G_f = \U(1)^f \times_{\phi} \text{p4}$, i.e. a triple $(A,\vec{R},\omega)$. $\vec{R}/2\pi \in \Z^2$ is a gauge field for the (magnetic) translation symmetry. Its flux physically corresponds to the Burgers vector of disclination and dislocation defects. A crucial observation is that in a system with $\MO$-fold rotation symmetry, the Burgers vectors $\vec{b}$ and $U(\frac{2\pi}{\MO})\vec{b}$ are equivalent up to a global rotation of coordinate axes, and differences of the form $(1-U(\frac{2\pi}{\MO})) \vec{b}$ are trivial fluxes of $\vec{R}$. This implies the following quantization condition on a closed manifold $\mathcal{M}^2$:
\begin{equation}
    \frac{1}{2\pi}\oint_{\mathcal{M}^2} d\vec{R} \in \left(1-U(\frac{2\pi}{\MO})\right)\Z^2~. 
\end{equation}
Following the derivation sketched in Sec.~\ref{sec:inv-tqft}, we obtain the action
\begin{align}\label{eq:mainresponse}
    \mathcal{L} &= \frac{C}{4\pi} A \wedge dA + \frac{\SO}{2\pi} A \wedge d\omega +\frac{\ell_{\OO}}{4\pi} \omega \wedge d\omega +\frac{\vec{\mathscr{P}}_{\OO}}{2\pi} \cdot A \wedge d\vec{R} \nonumber \\ & +\frac{\vec{\mathscr{P}}_{s,\OO}}{2\pi} \cdot \omega \wedge d\vec{R} +\frac{\kappa}{2\pi} A \wedge A_{XY} +\frac{\nu_s}{2\pi} \omega \wedge A_{XY}.
\end{align}
Here $A_{XY}$ is an area form: if $\vec{R} = (X,Y)$,  $A_{XY} = \frac{1}{2\pi}X \wedge Y$ when $\omega = 0$ (the definition is more complicated when $\omega \neq 0$~\cite{manjunath2021cgt}). 

While Eq.~\ref{eq:mainresponse} captures all the topological invariants at once, it can be useful to define $\SO$ and $\ell_{\OO}$ (equivalently, $\Theta^{\pm}_{\OO}$) for each rotation center $\OO$ separately and understand the coefficients in Eq.~\ref{eq:mainresponse} in terms of their linear combinations~\cite{manjunath2023characterization}. Most of the crystalline invariants predicted by Eq.~\ref{eq:mainresponse} can be understood in this way. The two that cannot be fully determined from $\Theta^{\pm}_{\OO}$ are the integer $\kappa = \nu-C\phi/2\pi$, and the angular momentum filling $\nu_s$. In the Wannier limit, these equal $n_{\alpha} + n_{\beta} + 2n_{\gamma}$ and $m_{\alpha} + m_{\beta} + 2m_{\gamma} \mod 4$.

\subsection{Quantized electric polarization}\label{sec:pol}

The most important new piece in Eq.~\ref{eq:mainresponse} is the invariant $\PO$, which is a many-body electric polarization and can be detected in terms of the excess charge at lattice \textit{dislocations}. Here $\PO$ is origin-dependent because it is the polarization of the electronic system alone, and does not include contributions from any neutralizing background~\cite{zhang2025pol}. The field theory predicts that the electric charge at a dislocation with Burgers vector $\vec{b}$ receives a quantized contribution $\PO \cdot \vec{b}$ mod 1 from this term. For this to be invariant under a global rotation of coordinate axes, we must have
\begin{equation}
    \left(1-U(\frac{2\pi}{\MO})\right)^T  \PO \cdot \vec{b} = 0 \mod 1~.
\end{equation}
When $\MO = 4$, and $U(\pi/2)$ is the standard $2\times 2$ rotation matrix $i \sigma^y$, the values of $\PO$ take the form $\frac{\overline{\mathscr{P}_{\OO}}}{2}(1,1)$ with $\overline{\mathscr{P}_{\OO}} \in \Z_2$. It can be shown that $\PO$ is a linear combination of $\SO$ for different $\OO$, and $\kappa$~\cite{zhang2022pol}; for instance,
\begin{equation}
    \overline{\mathscr{P}}_{\alpha} = \mathscr{S}_{\beta} - \mathscr{S}_{\alpha} + \kappa \mod 2.
\end{equation}
A coloring of Hofstadter's butterfly for $\PO$~\cite{zhang2022pol} is shown in Figure~\ref{Figure_1}(d). It is a major triumph of the TQFT approach that we can now define a many-body electric polarization in Chern insulators. The relation between this many-body definition and previous single-particle definitions of electric polarization in Chern insulators~\cite{coh2009} has also been clarified~\cite{vaidya2024pol,zhang2025pol}. Surprisingly, the electric polarization can also be extracted in Chern insulators from edge and  corner charges~\cite{zhang2025FCI}, belying the naive expectation that the gapless edge modes in Chern insulators might make such a definition impossible.

\subsection{Relation to free fermion classification and band invariants}\label{sec:freefermion}
As shown in Table~\ref{tab:classification}, the analogous free fermion classification (when $\phi = 0$) is given by the group $\Z^9$~\cite{Kruthoff2017bandCombinatorics,manjunath2023characterization}. One factor corresponds to $c_- = C$, while the other invariants are labelled by $\{n^{\bf p}_q\}$ and specify the number of bands transforming with angular momentum $q$ mod 2 or 4 at the momentum-space high symmetry points ${\bf p} = \{\Gamma,M,X\}$. While there are $4+4+2 = 10$ such integers, there are also two relations which require $\sum_q n^{\bf p}_q = \kappa$ for each ${\bf p}$, giving 8 independent integers apart from $C$. It is clear that the free-to-interacting map for this symmetry is not one-to-one: most of the $\Z$ factors get reduced to $\Z_m$'s, or trivialized altogether. The map is also not onto, since interacting systems allow for $c_- \neq C$. The precise map between band invariants and TQFT invariants (assuming $\phi = 0$) was derived in Reference~\citenum{manjunath2023characterization}; see Reference~\citenum{lee2026freeinteracting} for other recent progress.

\section{Crystalline invariants in fractional Chern insulators}\label{sec:fci}

In this section, we review some key recent developments on characterizing crystalline invariants in fractional Chern insulators by taking the 1/2 Laughlin topological order as our main example. The topological order consists of two topologically distinct charges, $\{[0], [1]\}$, where $[1]$ labels the semion with topological twist $\theta_{[1]} = e^{i\pi/2}$. We have the fusion rules $[a] \times [b] = [a + b]$, where the brackets imply mod $2$ reduction. 

Initially, assume the global symmetry is $G = \U(1) \times \mathbb{Z}_4$, where $\mathbb{Z}_4$ is a $4$-fold spatial rotational symmetry generated by the operator $\hat{C}_{4,l} \equiv e^{i \frac{2\pi l}{4} \hat{N}} \hat{C}_4$ for integer $l$, with $\hat{N}$ the total $U(1)$ number operator. While the quantities $s,k_2,k_3$ introduced below all have an origin dependence, we will suppress it in the next subsection for ease of notation. 
\subsubsection{Classification} \label{sec:fci-classif}

The symmetry fractionalization class can be specified by two anyons, $[v], [s] \in \{[0], [1]\}$~\cite{manjunath2021cgt}. $v$ is the charge vector (vison) while $s$ is a crystalline analog of the spin vector~\cite{Wen1992shift}. $v$ determines the Hall conductivity via $\sigma_H = \frac{1}{2\pi} (\frac{v^2}{2} + 2 k_1)$ in natural units, as well as the fractional charge of each anyon $a\in\{[0],[1]\}$, $Q_a = a v/2 \mod 1$. $s$ determines the discrete shift $\mathscr{S} = \left(\frac{vs}{2} + k_2\right)$ mod 4, which in the topologically ordered case also specifies the fractional orbital angular momentum of the anyons, $L_a = a s /2 \mod 1$. These invariants appear as coefficients in an effective CS theory:
\begin{align}
\label{effAc}
    \mathcal{L} = \frac{2}{4\pi} a d  a - \frac{v}{2\pi} A d  a - \frac{s}{2\pi} \omega d  a + \mathcal{L}_{\text{SPT}},
\end{align}
where $\mathcal{L}_{\text{SPT}} =  -(\frac{k_1}{2\pi} A d  A  + \frac{k_2}{2\pi} A d  \omega +  \frac{k_3}{2\pi} \omega d  \omega)$ for integers $k_1,k_2,k_3$. Here $a$ is a dynamical $U(1)$ gauge field, $A$ is the background $U(1)$ gauge field, and $\omega$ is a background $\mathbb{Z}_4$ gauge field of the crystalline symmetry. We also have equivalences $(s,k_2, k_3) \sim (s - 2, k_2+v, k_3+s - 1 )$ and $(v, k_1,k_2) \sim (v -2,k_1+v-1,k_2 + s-1)$, which arise by relabeling $a \rightarrow a + \omega$ and $a \rightarrow a + A$ respectively in Eq.~\ref{effAc}.

\subsubsection{Discrete torsion vector and fractionally quantized electric polarization}

While the spin vector has a known continuum analog~\cite{Wen1992shift}, including translation symmetry leads to a new form of crystalline symmetry fractionalization with no continuum analog: the \textit{discrete torsion vector} $\vec{t}_{\OO}$ ~\cite{manjunath2021cgt,Manjunath2020fqh} (here and below we restore origin dependences on the coefficients) which depends on both rotational and translational symmetry. This is defined by a new term $    \frac{\vec{t}_{\OO}}{2\pi} \vec{R} \cdot da$ that generalizes Eq.~\ref{effAc}. Arguments like in Section~\ref{sec:pol} imply that for an order 4 rotation center, $\vec{t}_{\OO} \simeq \vec{t}_{\OO} + (1,1)$, giving $\vec{t}_{\OO}$ a $\Z_2$ classification; the $\Z_2$ invariant is given by $t_{\OO} = t_{\OO,x} + t_{\OO,y}$ mod 2. Physically, the trivial choice corresponds to a picture where we effectively induce an integer number of anyons at any dislocation. The symmetry fractionalization is non-trivial when no such picture exists, and one effectively has a `fraction of an anyon' at a dislocation. In analogy with $s_{\OO}$ and the fractional angular momentum of anyons, it is tempting to use $\vec{t}_{\OO}$ to define a fractional linear momentum for each anyon, but it is an open question to make this notion microscopically precise. There are combined equivalences on $s_{\OO},\vec{t}_{\OO}$ coming from gauge field relabellings analogous to those below Eq.~\ref{effAc}.

$\vec{t}_{\OO}$ can be understood in terms of differences of $s_{\OO}$ using a real-space construction, in a similar spirit to the Wannier-limit constructions in invertible states: for example, on the square lattice, $t_{\OO} = s_{\OO} + s_{\gamma}$ mod 2 when $\OO = \alpha$ or $\beta$. ~\cite{kobayashi2024FCI}. Moreover, just as $s_{\OO}$ specifies a fractionally quantized discrete shift $\SO$, $\vec{t}_{\OO}$ specifies a fractionally quantized electric polarization $\PO$. When $\OO$ is an order 4 rotation center of the square lattice, $\PO = \frac{\overline{\mathscr{P}_{\OO}}}{2}(1,1)$, where $\overline{\mathscr{P}_{\OO}}= \left(\frac{v t_{\OO}}{2}+ k_{4,\OO}\right)$ mod 2. In particular, dislocations in the 1/2-Laughlin FCI can carry a minimal charge in multiples of $1/4$, even though the minimal anyon charge is $1/2$. 

Finally, there is an additional fractionalizaton of translation symmetry alone, given by an `area vector' $m$, which specifies an `anyon per unit cell' and sets the fractional part of the filling, $\nu = \frac{vm}{2}$ mod 1; this has been discussed in detail in prior work~\cite{sachdev1999translational,cheng2016lsm}.

\subsubsection{Characterization by partial rotations and defect response} Significantly, the methods of partial rotations and defect response discussed in Section~\ref{sec:invertible} can also be used to characterize FCIs, with minor modifications. To evaluate each partial rotation, we pick a rotationally symmetric subregion $D$ whose length $|\partial D| \gg \xi$, with $\xi$ the correlation length. The partial rotation 
satisfies an analog of Eq.~\ref{eq:mainresultCFT} with a quantized complex phase, from which
the topological invariants $\Theta_{l,\OO}$ are obtained by taking appropriate modular reductions~\cite{kobayashi2024FCI}. 
By computing $\Theta_{l,\OO}$ for different $l$ and $\OO$, together with $\sigma_H$, 
one can \textit{completely} determine the symmetry fractionalization data $v,s_{\OO},\vec{t}_{\OO}$ as well as the SPT contributions $k_i$ for the 1/2-Laughlin FCI. One can also generalize the charge response in Eq.~\ref{eq:QW-final} to the case of FCIs, with the caveat that the topologically invariant part of the charge $Q_W$ is $Q_W$ mod $q_{\star}$ where $q_\star$ is the minimal quasiparticle charge (which is $1/2$ in the 1/2-Laughlin FCI).

\subsubsection{Numerical Monte Carlo simulation}
We construct an ideal wave function for the 1/2 Laughlin FCI using the parton construction, placing each parton in a Hofstadter ground state wave function with magnetic flux $\phi_{p,i}$ per unit cell. For the projection to survive, each parton state must have the same filling $\nu_p =  \frac{\phi_{p,i}}{2\pi} + \kappa_i$, for $i = 1,2$. Each state has known invariants $\{\mathscr{S}_{\OO}, \ell_{\OO}, \dots \}$ in addition to $C$ and the filling $\nu$, and the construction completely determines the coefficients of Eq.~\ref{effAc} in terms of the invariants of the parton states~\cite{kobayashi2024FCI}. The quantization of each invariant predicted by the above classification has been verified in detail~\cite{kobayashi2024FCI,zhang2025FCI} through wave function Monte Carlo simulations.

\begin{summary}[SUMMARY]
\begin{enumerate}
    \item The approaches to classify and characterize a topological state depend on whether it is in a free-fermion, many-body invertible, or topologically ordered phase.
    \item The classification of invertible states is obtained by combining ideal wave function and TQFT/category-theoretic approaches.
    \item Invertible states are characterized using (i) defect responses (prominent examples are the discrete shift $\SO$ and electric polarization $\PO$) and (ii) partial symmetry observables $\Theta^{\pm}_{\OO}$, which are both deeply connected to TQFT.
    \item Partial symmetries can fully characterize the crystalline invariants in Chern insulators, given the Chern number $C$ and filling per unit cell $\nu$.
    \item In the topologically ordered case, the crystalline symmetry fractionalization data includes a spin vector $s_{\OO}$, a torsion vector $\vec{t}_{\OO}$ and an `anyon per unit cell' $m$. $s_{\OO}$ and $\vec{t}_{\OO}$ endow the anyons with fractional angular and linear momentum respectively, and result in fractionally quantized responses $\SO$ and $\PO$. These can also be determined through defect responses and partial symmetries. 
\end{enumerate}
\end{summary}

\begin{issues}[OUTLOOK]
\begin{enumerate}
\item What values do the invariants described here take in FCIs experimentally realized in moir\'e materials~\cite{xie2021fci,cai2023fci} or artificial quantum simulators~\cite{clark2020laughlin}? As described here, these invariants can be easily computed in model wave functions of Chern insulators, and subsequently deduced for simple FCIs using parton constructions. The topological invariants are encoded in the discrete shift, the electric polarization, and in partial rotation observables, potentially giving multiple routes for their measurement.
\item There is a wealth of open directions related to measuring these invariants experimentally; for instance, (i) detecting fractional charges through local probes in 2d materials; (ii) creating crystalline defects and measuring of analogous fractional charges in photonic topological systems, in which CIs and FCIs have been realized~\cite{hafezi2013,ozawa2019,clark2020laughlin}; (iii) Measuring partial rotations in quantum simulators. It is well-known that the expectation value of a partial SWAP operator can be measured using random unitaries, and in fact rotations of order 2 are just SWAPs; the question is now to consider general rotations and implement them on current experimental platforms. 
\item There are a number of physically interesting symmetries in which individual examples exist, but we are not aware of a complete classification. These include: (i) free-fermion systems with $\phi \neq 0$; (ii) magnetic translations with $\Z_N$ flux in invertible and SET states; and (iii) FCIs with anyon-permuting crystalline symmetries (the case of quantum spin liquids was studied recently~\cite{ye2024QSL}).
\item An important theoretical direction is to systematically include reflections and time-reversal in the $G$-crossed BTC framework. Some progress has been made on these lines~\cite{barkeshli2019tr,Barkeshli2020Anomaly} but the general theory is still incomplete.
\item There are some existing theoretical predictions that have not yet been numerically tested. For instance, `partial reflection' invariants for bosonic SPTs with wallpaper group symmetries have been proposed~\cite{calvera2025wallpaper}, but we are not aware of detailed numerical studies involving them. It is also an open question to numerically verify the partial rotation predictions from CFT in the case of fermionic FCIs. Note that the existing results for partial rotations are based on truncating the CFT expansion to leading order, and it would be interesting to investigate higher-order terms in the expansion.    
\item A particularly interesting frontier is the characterization of crystalline topological invariants in gapless systems. For example, (2+1)D Dirac fermions- a particularly simple gapless system- were recently shown to have the property that a disclination of the UV rotation operator induces an `emanant' magnetic flux in the low-energy theory~\cite{barkeshli2026disclinations}, with several observable consequences such as an azimuthal current. 
\end{enumerate}
\end{issues}

\section*{DISCLOSURE STATEMENT}
The authors are not aware of any affiliations, memberships, funding, or financial holdings that
might be perceived as affecting the objectivity of this review. 

\section*{ACKNOWLEDGMENTS}
We thank Yuxuan Zhang, Gautam Nambiar, Vladimir Calvera and Ryohei Kobayashi for highly rewarding collaborations spanning several years. Research at Perimeter Institute is supported in part by the Government of Canada through the Department of Innovation, Science and Economic Development and by the Province of Ontario through the Ministry of Colleges and Universities. MB is supported by NSF DMR-2345644.

\bibliographystyle{ar-style4}  
\bibliography{References,refs}

\end{document}